# A dual set-up based on Bradley's aberration of light, using simultaneously stellar and local light sources.

G. Sardin
gsardin@ub.edu

## Abstract

A dual optical set-up is proposed to detect simultaneously the different behavior of light from stellar and local sources, in relation to speed-induced aberration. A small laser is set at the center of the objective lens of a telescope, allowing to record at once the two spots on an array detector. Their positions are recorded during a yearly earth orbit, so stellar aberration can be visualized as a tiny circle. But no aberration has been observed from local sources, hence the laser spot should remain still. The simultaneous recording of both spots allows highlighting their different behavior. Einstein related aberration to the transverse speed between light source and observer, and since for local sources it is null, no aberration ensues. Despite this explanation conforms with the correct result for local sources, it cannot however be retained since stellar aberration does not vary although relative speed differs for each star. Consequently, the null transverse velocity cannot be considered the cause of the null aberration from local sources. A causal approach to this different behavior between stellar and local light is advanced, based on the combined effects of a speed-induced deflection of emitted light rays and a speed-induced aberration upon detection.

## 1. Introduction

Already about 1680 the french astronomer J. Picard observed that the position of star was not steady but underwent small shifts, feature verified by the english astronomer J. Flamsteed. In 1725, Bradley (1) carefully determined that in fact the observed position of stars undergo a yearly small rotation. He deduced that it could not be attributed to the parallax effect and found out its actual cause, correctly interpreting it as an aberration due to the earth speed around the sun, now referred as the stellar aberration of light. Bradley established in its article of 1728 that the aberration angle is constant for all stars and expressed it as:

tn $\alpha$ = v / c      ( where v is the earth orbital speed and c that of light).

However, when using light from a local source, i.e. located on earth, instead of that from stars, no aberration has been observed (2-5). It is commonly trusted that special relativity has solved the absence of aberration from local sources, in view that Einstein linked aberration to the relative speed between the light source and the observer (6-10). Since for sources on earth the relative speed is null no speed-induced aberration is thus expected. However, despite this explanation complies with the actual outcome for local sources, it should be discarded since stellar aberration has been experimentally set to be constant for all stars and thus independent of their different speed relative to us, resulting to be determined only by the earth speed around the sun (11-17). Therefore, the fact that for local sources the relative speed is null cannot coherently be considered the accurate reason for the null aberration and some other causal clarification should be searched (18).

The device proposed takes advantage of the presence of aberration of light from stars and of its absence from local sources, in order to ratify it and to allow improving the causal insight of this different behavior of light depending on the source type, stellar or local. Indeed, light rays from stellar sources undergo the Bradley effect, i.e. the aberration of stellar light which leads to a mirage position of stars due to an apparent inclination of their rays caused by the earth speed. However, since no aberration is detected from local sources, it can be expected that the rays from these two different sources will undergo, in a year, paths gradually deviating from each other. Based on this oddity, a specific set-up is broached in order to experimentally evidence these different behaviors and at the time to arouse the search for solving the speed-induced aberration dilemma between stellar and local sources.

**2. Experimental:** description and working way of the dual set-up

The working way of the device proposed is specifically based on the behavior difference of rays from stellar and local light sources. Since one leads to an aberration and the other one does not, a simultaneous differential measure of the two cases should provide a conducive contrasted result. In effect, one spot should stay immobile on the array-detector while the second should be mobile in accordance to the earth motion around the sun. So, an hybrid set-up based on the simultaneous detection of light rays from a star and from a laser has been devised. The system is made of a telescope on whose objective is added a laser fixed on its center. At the telescope focal a high resolution photo-detector array is set, allowing to simultaneously track the position of the spots from the star and the laser. If the telescope is pointed e.g. toward the polar star, in the course of one year the apparent position of the star will describe a quasi circular ellipse. However, since no aberration has been detected from any local source it is thus expected that on its part, the spot from the laser beam placed on the telescope objective will remain still on the array-detector.

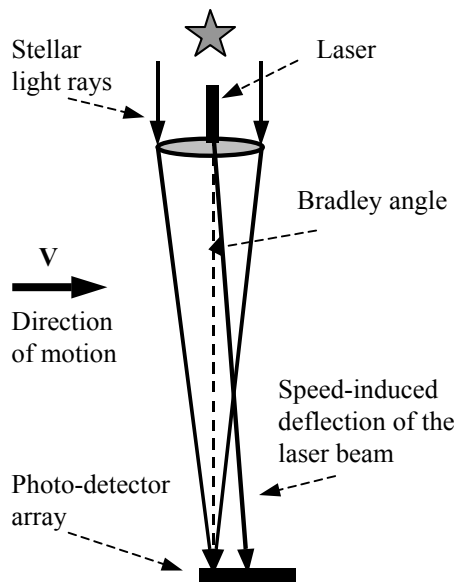

**Fig.1**: Hybrid Bradley speed-meter, based on the simultaneous detection of the spot position on a photo-detector array of the impinging light rays from stellar and local (laser beam) sources.

The working procedure sets in pointing the laser beam such as its spot on the array detector superposes to that from the star and to daily record at the same sideral hour their position during a year, in order to clearly display the temporal difference of behavior of the two spots. In effect, within the telescope, during the time-of-flight of the light rays from the objective to the focus, the detector array has slightly moved sideways from its position in space at the very moment the rays were impinging on the objective. This creates an apparent tilt of the incoming light rays leading to a mirage position of stars, whose angle is expressed by tn $\alpha$ = v / c and known as Bradley constant (corresponding to an aberration angle of 22.4"). For a telescope with a focal length (f) of 1 m the time of flight of photons is:

$t = f / c = 1 / c = 1 / 3.10^{-8}$ m/s

and the earth speed being of 30 km/s the meanwhile lateral shift of the detector is:

$\triangle x = v.t = (3.10^4)(1/3.10^{-8}) = 10^{-4}$ m = 100 μm

Equivalently, dx can be directly formulated as:

$\triangle x = f.tn(v/c) = 1. tn(30 / 300000) = tn (10^{-4}) \approx 10^{-4}$ m = 100 μm

So, after one year the spot from the stellar light should have drawn a tiny circle of a 100 μm radius, while the spot from the local source should have remained still. Already after six months the earth speed vector is inverted and hence the shift from the stellar spot is inverted too. In contrast, the laser spot remains immobile since the beam undergoes a speed-induced deflection, which is always oriented in the direction of motion. Let us now apply the speed-induced aberration upon detection and contrast it with the speed-induced actual deflection of light rays.

### 3. Discussion

Contrasting the different results from the stellar and the local sources should help developing an adequate causal interpretation of the Bradley effect for both cases, since it has been previously expressed that the relative speed between source and observer does not manage to differentiate suitably the two cases. In effect, it has already been highlighted in the introduction that the relative speed between source and observer cannot be taken as an accurate cause for the Bradley aberration. In its place, an alternative is advanced, based instead on the resultant of two effects, one proceeding from the source motion and the other one from the observer motion. It appears necessary to consider these two factors in order to provide a congruent causal interpretation of the difference of behavior between stellar and local sources.

### 3.1. Speed-induced deflection of light rays upon emission

The deflection of emitted light-rays induced by the source speed is considered an actual physical effect, the rays path being truly deflected and it should not be mistakenly taken for the mirage deflection induced by the observer speed. The equivalency of inertial systems implies that the spot on the array must stay still under any change of motion of the inertial system. This requirement is fulfilled only if the component $c_x$ of the speed of light in the direction of motion is equal to the system speed v (fig.1). This implies the reflected beam to be deflected by an angle $\alpha$ such that:

$c_x = c . \sin \beta = v$

Let us suggest causal grounds for the assumed speed-induced deflection. The deflection angle is equal to:

$\beta = \arcsin (c_x / c) = \arcsin (v / c)$

It appears as a function of the ratio v / c, but this relation may be regarded as the reduced result of the energy ratio $pc / mc^2$. Hence $\sin \alpha$ can be rewritten in terms of an energy ratio:

$\sin \beta = E_r / E_i = pc / mc^2 = mvc / mc^2 = v / c$

where $E_r = pc = h\nu$ and $E_i = mc^2 = h\nu_0$. $E_r$ stands for the relative collision energy, i.e. that measured by the detector of the inertial system, and $E_i$ for the intrinsic collision energy, i.e. that a detector at absolute rest would measure. So, the deflection angle can be looked at due to a transverse momentum of the emitted photons, i.e. a momentum in the direction of the source motion.

### 3.2. Speed-induced aberration of light rays upon reception

Let us now attend the cause of the proper aberration. The aberration of light rays relative to the observer is due to its own motion, and being an apparent effect the rays are in fact not truly deflected but they appear so to the observer. The speed-induced aberration of light rays ensues from the finite speed of light, which implies a finite time-of-flight of light from the objective to the eye-piece or detector. This leads to an apparent deflection equal to:

$$\tn \alpha = L_x / L_y = v \cdot t / c \cdot t = v / c$$

A simple way to depict the exclusive local nature of stellar aberration could be by considering a long tube with pin-hole diaphragms at its ends. Light rays parallel to the tube axis passing across the top pin-hole diaphragm will not come out from the tube if it has moved tangentially during the rays time-of-flight, due for e.g. to the earth motion. To highlight this point let us cite the pertinent quotation of C. Renshow (19): "Aberration is therefore clearly a local phenomenon. Light from a distant star could be heading toward Earth for a billion years, say from due north. As it is viewed on Earth, it will be aberrated one way or another, depending on the particular season it reaches us. Due to extreme distance of the star, aberration will cause the apparent line-of-sight to betray the star's "true" position by hundreds to thousands of light years. The source is certainly not jumping around like this, and the light must be coming directly from the actual source location to the Earth along the line joining the location of the star and Earth, not along the aberrated line of sight".

Let us now outfit causal grounds for the presence of aberration from stellar light and of its detectionless in the case of local light, on the basis of the adjoined contribution of the speed-induced deflection and aberration. As already stated the presence of aberration is considered due to two speed-induced effects, i.e. speed-induced deflection and speed-induced aberration. Let us analyze their net outcome for stellar and local light sources.

### 3.3. Cause of stellar aberration

It has been assumed that speed induces a deflection of light rays in the direction of motion whose angle $\beta = \arcsin(v/c)$, but the emission of light from star has a spherical symmetry and thus the deflection of the rays cannot be discerned since the net effect is then null. This leads to an independence from the effect of speed on stellar light and the expected result is the same as if it is considered that there is no speed-induced deflection of light-rays, just like if the star would be at absolute rest. In view that the spherical symmetry leads to the same result in both cases, deflection or deflectionless, why then any need in assuming a speed-induced deflection since it seems useless? Its usefulness emerges when considering the absence of aberration from local sources. In any case, the effect of speed upon the direction of emission of light rays from sources with spherical symmetry is not observable and so does not affect the observed aberration from star.

It ends up therefore that the observed aberration from star light is determined exclusively by the effect of the observer motion on the apparent direction of the oncoming rays upon detection. This observational effect leads to a fictitious inclination of light rays due to the observer motion, which are actually impinging parallel to the telescope axis. This apparent tilt ($\tn \alpha = v/c$) of light rays is determined by the objective position in space at the instant the rays impinge on it and the ocular position at the time they reach it. Equivalently, in terms of drift ($\Delta x = d \cdot \tn \alpha$), the aberration derives from the shift between the ocular position in space at the very moment the rays impinge on the objective, and the ocular new position at the time the rays reach it.

### 3.4. Cause of local aberrationless

Let us consider now the case of the light rays from the laser set on the center of the objective lens. In this specific case the emitted rays have no spherical symmetry but are instead directional and embodied into a beam whose path is affected by the system speed and tilted in the direction of motion of an angle $\beta$ = arc sin(v/c). It ensues that the emitted rays, being slightly tilted in the direction of motion, impinge on the telescope with an angle of incidence equal to that of the Bradley aberration and hence aberration cannot be observed since the two effects offset. This is due to the time of flight (t) of photons to cover the length from the objective to the ocular, and so the aberration ($\triangle x = c_x.t$) is exactly balanced by the meanwhile corresponding dislodge ($\triangle x = v.t$) of the eye-piece or detector, so rays impinge at the very same point on the photodetector array (fig.1).

**Conclusion**

The proposed dual arrangement allows to contrast the different behavior of light from stars or from local sources in regard to the Bradley effect. It has been highlighted that in spite of the fact that effectively no aberration has been observed from local sources, the apparently seemly explanation formulated by Einstein which relates aberration to the relative speed between the light source and the observer cannot be retained although it conforms with the null observation in the case of local sources, since both being on earth their velocity relative to the observer is then null and thus no aberration should consequently be expected. But, on another part it has been experimentally settled that stellar aberration is independent of the relative speed between star and observer since whatever the star observed, the aberration remains constant in spite of the different relative speed in each case. Therefore, despite the null relative speed between local source and observer this cannot logically be considered the cause of the absence of aberration.

This remark calls for a different cause to explain the different behavior of light from stars and from local sources, other than the relative speed. The interpretation proposed stands instead on the relative speed ($c_x$) between the component of the speed of light along the direction of motion and the speed (v) of the optical system. When both are equal ($c_x$ = v) there is no aberration of light. In the case of stellar light, $c_x$ is null since the rays are then impinging parallel to the optical axis, and the aberration angle is thus only dependent on v. In the case of local sources the speed induced deflection of light rays is equal to the Bradley angle and thus the relative speed $\triangle v$ is then null ($\triangle v$ = $c_x$ - v = 0).

So, the described set-up aims to bring a contrasted experimental proof of the different behavior of light according to its origin, stellar or local. In both cases, their light rays impinge on the same objective and their traces on the same array detector are simultaneously recorded, allowing to detect any variation of their position, during e.g. the yearly cycle of earth around the sun. The linked conceptual aim of the experiment stays in providing crucial clues in order to apprehend the factors determining the difference of behavior of light from stellar and local sources. The cause for the difference has not been deeply analyzed due to the common belief that Einstein theory of relativity has provided the answer. An alternative causal explanation has been proposed, based on the combined effects of two factors: the speed-induced deflection of light rays from moving sources, which is an effective effect, and the speed-induced apparent deflection of the light rays path through observation, which is an optical mirage induced by the observer speed. In the case of local sources both effects compensate and the net result is null, due to the fact that rays impinge on the objective with an incident angle equal to the aberration angle.

**References**

1. J. Bradley, Phil. Trans. **35** (1728)
2. L. Respighi, Mem. Accad. Sci., 2 (1861)
3. M. Hoek, Astr. Nach., 73 (1868)

4. E. T. Whittaker, *A History of the Theories of Aether and Electricity* (Longman Green & Co, London, 1910)
5. T. Theocharis, Speculation in Science and Technology, 15, 1 (1992)
6. R. Resnick, *Introduction to Special Relativity* (John Wiley & Sons, Inc., New York ,1968)
7. A. French, *Special Relativity: The M.I.T. Introductory Physics Series* (W. W. Norton, New York, 1968)
8. W. K. Panofsky and M.Phillips, *Classical Electricity and Magnetism* (Addison-Wesley, 1955)
9. J.P. Cedarholm, C.H. Townes, Nature, **184** (1959)
10. T. Jaseva, A. Javan, J. Murray, C. Townes, Physical Review **133**, A (1964)
11. P. Naur, Physics Essays 12, 2 (1999)
12. E. Eisner, Am. J. Phys. **35** (1967)
13. T.E. Phipps, Am. J. Phys. **57** (1989)
14. H.E. Yves, J. Opt. Soc. Am. **40** (1950)
15. H. C. Hayden, Galilean Electrodynamics **4** (1993)
16. P. Marmet, Physics Essays **9**, 1 (1996)
17. D. Johnson, Galilean Electrodynamics (to be published)
18. G. Sardin, Europhys. Lett., **53**, 3 (2001)
19. C. Renshaw, IEEE Aerospace Conf. (Snowmass, Colorado, March 1999)

*Author's other related articles:*

First and second order electromagnetic equivalency of inertial systems, based on the wavelength and the period as speed-dependant units of length and time

A causal approach to first-order optical equivalency of inertial systems, by means of a beam-pointing test-experiment based on speed-induced deflection of light.

lanl.arXiv.org e-Print archive (url: xxx.lanl.gov),  Physics, Subj-class: General Physics